\documentclass[conference]{IEEEtran}
\IEEEoverridecommandlockouts
\usepackage{cite}
\usepackage{amsmath,amssymb,amsfonts}
\usepackage{algorithmic}
\usepackage{graphicx}
\usepackage{textcomp}
\usepackage{xcolor}
\usepackage{hyperref}
\usepackage{braket}
\def\BibTeX{{\rm B\kern-.05em{\sc i\kern-.025em b}\kern-.08em
    T\kern-.1667em\lower.7ex\hbox{E}\kern-.125emX}}

\makeatletter
\newcommand{\linebreakand}{%
  \end{@IEEEauthorhalign}
  \hfill\mbox{}\par
  \mbox{}\hfill\begin{@IEEEauthorhalign}
}
\makeatother

\begin{document}

\newcommand{\ignore}[1]{}
\newcommand{\sid}{\textcolor{red}}
\newcommand{\tina}{\textcolor{orange}}
\newcommand{\revision}{\textcolor{blue}}

\title{TUSQ: Tracking, Uncomputation, and Sampling for Noisy Quantum Simulation}

\author{
\IEEEauthorblockN{Siddharth Dangwal\textsuperscript{*}}
\IEEEauthorblockA{
\textit{University of Chicago}\\
Chicago, USA \\
siddharthdangwal@uchicago.edu}
\and

\IEEEauthorblockN{Tina Oberoi\textsuperscript{*}}
\IEEEauthorblockA{
\textit{University of Chicago}\\
Chicago, USA \\
toberoi@uchicago.edu}
\and

\IEEEauthorblockN{Ajay Sailopal\textsuperscript{*}}
\IEEEauthorblockA{
\textit{University of Chicago}\\
Chicago, USA \\
ajays@uchicago.edu}
\linebreakand
\IEEEauthorblockN{Dhirpal Shah}
\IEEEauthorblockA{
\textit{University of Chicago}\\
Chicago, USA \\
dhirpalshah@uchicago.edu}
\and

\IEEEauthorblockN{Frederic T. Chong}
\IEEEauthorblockA{
\textit{University of Chicago}\\
Chicago, USA \\
chong@cs.uchicago.edu}

\thanks{\textsuperscript{*}Equal contribution}
}


\maketitle

\begin{abstract}

Quantum computers have improved in size and quality in recent years, enabling the execution of complex circuits. However, for most researchers, access to compute time is limited. This necessitates the development of simulators that mimic noisy quantum hardware accurately and scalably. The ideal way to simulate noisy systems is via Density Matrix Simulation (DMS). However, its high memory footprint limits its scalability. Consequently, noisy simulations are performed in two steps: (a) sampling multiple circuits with fixed noisy gates from the stochastic noise channels, (b) performing the State Vector Simulations (SVS) of these circuits and averaging their output to obtain the effective noisy simulation result. This often leads to a substantial increase in compute overhead, slowing down the simulation. Existing methods solve this problem by caching critical intermediate results in memory and reusing them. However, when a simulation task is both compute and memory-intensive, we need to eliminate computational overheads without incurring extra memory overheads. 

To enable fast simulation in the compute and memory bound regime, we propose \textit{TUSQ} - \underline{T}racking, \underline{U}ncomputation, and \underline{S}ampling for Noisy \underline{Q}uantum Simulation. TUSQ is composed of two modules: the \textit{Error Characterization Module} (ECM), and \textit{Depth First Tree Traversal} (DFTT). The ECM characterizes errors so that the simulator can eliminate redundant circuit instances (via ER Tallying and ER Commutation), followed by importance sampling (in the Pruning stage), significantly reducing the number of circuits to be simulated relative to the baseline strategy of simulating all circuits. This is followed by DFTT, which computes the statevectors for these sampled circuits efficiently by taking advantage of
circuit similarity, representing similar circuits in a tree and using computation
and uncomputation to traverse the tree efficiently. TUSQ is evaluated for a total of 198 benchmarks, executed for 1 million shots and reports an average speedup of $59.06\times$ and $13.38\times$ over Qiskit and CUDA-Q, with a maximum speedup of $7878.03\times$ and $439.38\times$ respectively. We also compare TUSQ against TQSim in the time and memory critical regime. We observe an average and maximum speedup of $39.32\times$ and $3134.31\times$, respectively.

\end{abstract}

\begin{IEEEkeywords}
Noisy Quantum Simulation, Statevector Simulation, Tensor Network Simulation 
\end{IEEEkeywords}

\section{Introduction}\label{sec:introduction}


\begin{figure}
    \centering
    \includegraphics[width=0.9\linewidth]{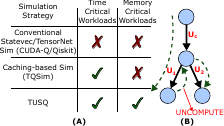}
    \caption{(A) Simulation tasks can be memory critical, time critical or both. For computationally cheap tasks, a conventional simulator like Qiskit GPU/CUDA-Q \cite{cudaq_noisy_sim, qiskit_gpu_statevec} works best since we avoid pre-processing time (which is usually performed on a CPU). For computationally expensive simulations that are not memory critical, we can cache critical intermediate results and reuse them to reduce simulation time. This is done in TQSim \cite{wang2025accelerating}. For tasks that are both time and memory critical, TUSQ eliminates computational redundancies using an initial pre-processing step. (B) For circuits with high overlap, TUSQ computes the output of the first, uncomputes to an intermediate stage and then carries out the remaining computation for the other, thereby reusing the computation for the initial overlapping portion.}
    \label{fig:intro_fig}
\end{figure}

Quantum computing has the potential to speed up tasks like factoring \cite{Shor_1997}, unordered search \cite{grover1996fast}, and physics and chemistry simulations \cite{kandala2017hardware, peruzzo2014variational}. However, compute time on these devices is scarce and expensive. It is often available only via cloud vendors like IBM \cite{IBMQE}, Quantinuum \cite{quantinuum_hardware}, and Quera \cite{quera_hardware} with long queue times \cite{ravi2021quantum}. To accelerate quantum computing research, we need simulators that accurately mimic the execution of programs on real, noisy quantum hardware. We place a special emphasis on noisy simulations. Most current quantum simulators focus on noiseless simulation, which is not representative of real hardware \cite{wu2019full, jiang2025bqsim, li2020density}. Hence, scalable, noisy simulators are the need of the hour.

\begin{figure*}
    \centering
    \includegraphics[width=\linewidth]{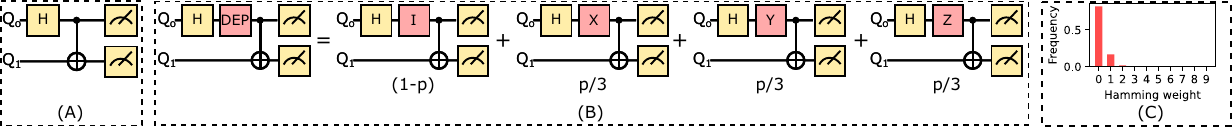}
    \caption{(A) A noiseless quantum circuit composed of one single- and one two-qubit gate. (B) A noisy version of the quantum circuit (A) with a non-unitary, stochastic depolarizing channel (DEP). This circuit can effectively be seen as a weighted classical average of four quantum circuits with the weights mentioned. Here, `p' is the probability of noise on the H gate and is bounded between 0 and 1. (C)Frequency of sampled ERs by hamming weight. ERs with lower hamming weight (more $I$ gates) are more frequent}
    \label{fig:qc_ideal_noisy_bg}
\end{figure*}

The most popular quantum computing paradigm is gate-based quantum computing \cite{nielsen2010quantum}. Noiseless gate-based systems can be simulated by starting in a known quantum state (represented as a complex vector), multiplying it with fixed unitary matrices (called gates) to obtain the final state and sampling it for a finite number of shots to output a classical probability distribution. A larger number of shots better represents the final quantum state, due to reduced sampling error. This method of performing noiseless \textit{Quantum Circuit Simulation} (QCS) is called \textit{State Vector Simulation} (SVS). Note that the matrix-vector multiplications needed to obtain the final vector need to be performed only once here. 

Noisy QCS, on the other hand, involves stochastic noisy operations, which may manifest as different gates every shot. Figure \ref{fig:qc_ideal_noisy_bg} (B) gives an example of one such operation (DEP), which can manifest itself as four different operations ($I/X/Y/Z$) with varying probabilities. To accurately account for this stochasticity, we need to perform the entire sequence of matrix-vector multiplications afresh every time we sample the output vector. Hence, if our output distribution consists of $S$ samples, we need to perform $S$ distinct SVSs, resulting in an $S$-fold compute overhead.

An alternative approach to noisy QCS is \textit{Density Matrix Simulation} (DMS), which can account for the effect of noise on a quantum state in one circuit execution. For this, DMS represents quantum states as matrices. Application of quantum operations is represented as matrix-matrix multiplication. Although DMS requires just one circuit execution, it has a quadratically higher memory overhead - $\mathcal{O}(2^{2n}$ as opposed to $\mathcal{O}(2^{n})$) in the case of SVS, which makes it an infeasible choice \cite{wang2025accelerating, patti2025augmenting, chen2021low, jones2019quest}. Ref \cite{wang2025accelerating} reports an estimate where El Capitan, one of the world's most powerful supercomputers, can handle at most a 25-qubit DMS, whereas one can run a 30-qubit SVS on a personal laptop with 16GB memory.

The excessive memory overhead of DMS makes the use of multiple SVSs the most efficient strategy for conducting noisy QCS. The accuracy of SVS for noisy QCS depends on the number of SVS samples ($S$) used. The statistical error w.r.t. DMS scales as $\epsilon = \frac{\sigma}{\sqrt{S}}$, where $\sigma$ is the standard deviation of the observable across the ensemble. For larger circuits (more gates), the value of $\sigma$ is also greater, which requires a bigger value of $S$ to keep $\epsilon$ low \cite{wang2025accelerating}. Further, more noisy gates imply a lower output fidelity, which also warrants an increased number of samples to get a non-zero useful signal. This, coupled with a longer time to perform each individual SVS, results in a rapid increase in the compute overhead of noisy QCS. Such workloads that take a long time to run on available hardware are called \textit{time-critical} workloads.

Recent noisy simulators like TQSim \cite{wang2025accelerating} handle this large compute requirement by caching critical intermediate results in memory and reusing them. This approach is quite practical when the workload has fewer qubits or the available memory is large. For example, TQSim simulates up to 20 qubit benchmarks (8MB memory consumption) on CPUs with 192 GB memory and on GPUs with 40GB memory. The additional memory allows caching of greater than or equal to 24576 and 5120 intermediate statevectors, respectively, which leads to a massive reduction in compute overhead. However, consider the simulation of a 30-qubit program on a 40GB GPU. The memory footprint of an $n$-qubit statevector scales as $\mathcal{O}(2^{n})$. A 30 qubit statevector has a size of $\sim 8GB$! This means a maximum of 5 statevectors can be cached at any given time, severely limiting our ability to reuse computation. Such workloads that consume the majority of available memory are called \textit{memory-critical}. The number of gates in a quantum program is often a function of the number of qubits. Thus, the exponential scaling of the statevector memory with the number of qubits, coupled with the exponential scaling of intermediate states in the number of gates (not all of them are critical though), makes a caching-based strategy sub-optimal in the time and memory critical regime.

This work addresses both compute and memory constraints by proposing \textit{TUSQ}: \underline{T}racking, \underline{U}ncomputation, and \underline{S}ampling for Noisy \underline{Q}uantum Simulation. Figure \ref{fig:intro_fig} (A) shows TUSQ's regime of interest. If the simulation is not compute-intensive, paying a CPU pre-processing cost is sub-optimal and a trivial simulation of the SVS ensemble on GPUs, as performed by Qiskit-GPU \cite{qiskit_gpu_statevec} and CUDA-Q \cite{bayraktar2023cuquantum}, is preferred. A caching-based solution like TQSim is preferred for computationally intensive tasks when additional memory is available. In the compute and memory-constrained regime, TUSQ's on-CPU pre-processing step for redundancy elimination in the SVS ensemble makes noisy QCS faster than other approaches.

TUSQ's pre-processing step comprises two modules - \textit{Error Characterization Module} (ECM), and \textit{Depth First Tree Traversal} (DFTT). Central to both modules is a lightweight Intermediate Representation (IR) called \textit{Error Realization} (ER) that enables us to determine the amount of computational redundancy among circuits. Performing matrix-vector multiplications to obtain the final statevector is much more expensive than sampling it multiple times \cite{patti2025augmenting}. The ECM uses ERs to identify circuits with the same final statevector, enabling their collective simulation by computing the common statevector once and sampling repeatedly from it. Once we are left with only unique circuits, the ERs also allow us to weigh the significance of each unique circuit and eliminate those with an insignificant contribution to the output. After this, DFTT computes the statevectors for these sampled circuits efficiently by taking advantage of circuit similarity, representing similar circuits in a tree and using computation and uncomputation to traverse the tree efficiently. Figure \ref{fig:intro_fig} (B) shows the simulation of two distinct circuits (represented by unitaries $U_1 U_c$ and $U_2 U_c$) with the same first half. TUSQ obtains the final statevector of the first circuit, uncomputes to the common intermediate stage and then carries out the remaining computation for the other, thereby saving on the time needed to simulate the common first half again. When generalized for all gates across all circuits, we effectively perform a Depth-First Tree Traversal (DFTT).

The net result of these optimizations is that we can simulate a 30-qubit adder circuit for $10^6$ shots on a 40GB Nvidia A100 GPU in $\sim 820$ seconds. The same benchmark on popular simulators like CUDA-Q and Qiskit takes more than 10 hours (keeping hardware specifications the same). Thus, TUSQ achieves high speed in the compute and memory critical regime. Overall, the key contributions of our work are:

\begin{enumerate}
    \item The \textit{Error Characterization Module} (ECM) eliminates all circuit instances which do not produce a distinct statevector or which contribute insignificantly (based on user-defined threshold parameters) to the final output. 
    \item \textit{Depth First Tree Traversal} (DFTT) groups circuits together based on the extent of their initial overlapping gates. The collection of circuits is represented as a tree, with the initial overlapping portion of two distinct circuits represented using the same edges, facilitating computational reuse.
    \item Although most of the discussion in the paper assumes Statevector simulations, TUSQ is backend agnostic. We create a GPU-compatible software implementation of TUSQ and evaluate noisy Statevector and Tensor Network simulations. An open-source implementation of TUSQ can be found \href{https://github.com/tinaoberoi/TUSQ}{\underline{here}}.
    \item We evaluate TUSQ on 198 benchmarks and report average speedups of $59.06\times$ and $13.38\times$ over Qiskit and CUDA-Q, with maximum speedups of $7878.03\times$ and $439.38\times$, respectively. We also compare TUSQ against TQSim \cite{wang2025accelerating} in the time and memory critical regime and observe an average and maximum speedup of $39.32\times$ and $3134.31\times$ respectively.  
\end{enumerate}

\section{Noisy Quantum Circuit Simulation}\label{sec:noisy_qcs}

In Section \ref{sec:introduction}, we described Statevector Simulation (SVS). Equation \eqref{eq:matrix_vector_multiplication} shows the vector representation of a quantum state ($\ket{\psi}$), and how it is manipulated using a unitary matrix, $U$, in the statevector formalism.

\begin{equation}
    \ket{\psi '} = U\ket{\psi} = \scriptsize
    \begin{pmatrix}
        a_{00} & a_{01} & \dots & a_{0,2^{k}-1} \\
        a_{10} & a_{11} & \dots & a_{1,2^{k}-1} \\
        \vdots & \vdots & \ddots & \vdots \\
        a_{2^{k}-1,0} & a_{2^{k}-1,1} & \dots & a_{2^{k}-1,2^{k}-1}
    \end{pmatrix}
    \begin{pmatrix}
        v_{0} \\
        v_{1} \\
        \vdots \\
        v_{2^{n}-1}
    \end{pmatrix}
    \label{eq:matrix_vector_multiplication}
\end{equation}

The statevector formalism is sufficient when representing noiseless systems. However, in the presence of noise, we rely on a different formalism called the density matrix formalism. Here, qubits are represented as a positive semi-definite Hermitian matrix $\rho$ with a unit trace \cite{nielsen2001quantum} called the density-matrix. Operations on the qubits are represented as maps between the initial state $\rho$ and final state $\rho'$ according to the equation $\rho'=\sum_{i}K_{i}\rho K_{i}^{\dagger}$. All quantum operations, noisy or noiseless, can be represented using DMS. For example, the application of a noiseless gate $U$, as given in Equation~\eqref{eq:matrix_vector_multiplication}, becomes $\rho'=U\rho U^{\dagger}$ in the density matrix formalism. Similarly, a depolarizing error channel acts as $\rho'=(1-p)\rho + \frac{p}{3}X\rho X + \frac{p}{3} Y\rho Y + \frac{p}{3} Z\rho Z$. Here, $p$ denotes the probability of noise corrupting the state $\rho$. Performing QCS by implementing the equations of the density matrix formalism is called density matrix simulation (DMS). DMS gives us an elegant way to perform noisy QCS. However, as mentioned in Section \ref{sec:introduction}, the memory overhead of DMS grows quadratically faster compared to SVS, which limits the number of qubits that can be simulated even on supercomputing clusters \cite{wang2025accelerating}. Note that the depolarizing channel output is a sum of four terms corresponding to four outcomes - the quantum state stays unchanged with probability $1-p$ and it is acted upon by $X/Y/Z$ gates, each with probability $\frac{p}{3}$ (see Figure \ref{fig:qc_ideal_noisy_bg} (B)). This output can be interpreted as a weighted classical mixture of the input state $\rho$ acted upon by unitary gates $I, X, Y $ and $Z$. This interpretation lets us represent the output of noisy QCS as a sum of the outputs of multiple SVSs, each with a fixed manifestation of the noisy channels. This gives us a way to use the state vector formalism in the presence of noise with some extra circuit overhead as shown in Figure \ref{fig:qc_ideal_noisy_bg}(B). Simulators like CUDA-Q \cite{patti2025augmenting, bayraktar2023cuquantum}, and Qiskit \cite{qiskit_gpu_statevec} perform noisy simulation by executing these circuits sequentially. \ignore{(in our case, the depolarizing channel is substituted by one of its constituent noisy gates - $I/X/Y/Z$)}

\begin{figure}
    \centering
    \includegraphics[width=0.75\linewidth]{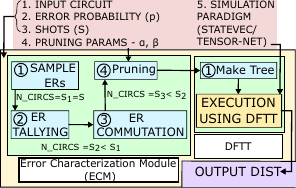}
    \caption{TUSQ Overview: (A) Error Characterization Module (ECM) - The ECM characterizes errors so that the simulator can eliminate redundant circuit instances (via ER Tallying and ER Commutation) followed by importance sampling (in the Pruning stage) significantly reducing the numbers of circuits to be simulated ($S_{3} \ll S_{1}$) (See Section \ref{subsec:ecm}) (B) DFTT then computes the statevectors for these sampled circuits efficiently by taking advantage of circuit similarity, representing similar circuits in a tree and using computation
    and uncomputation to traverse the tree efficiently (See Section \ref{subsec:dftt}).The inputs to TUSQ are highlighted in red above the figure. The final output distribution is highlighted in purple.}
    \label{fig:tusq_overview}
\end{figure}

\begin{figure*}
    \centering
    \includegraphics[width=0.9\linewidth]{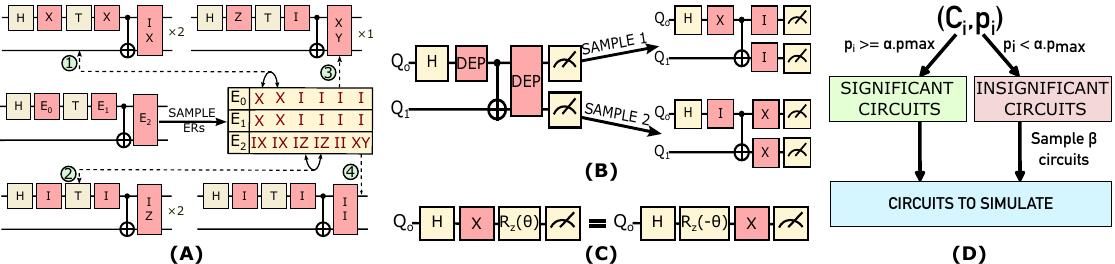}
    \caption{(A) ER tallying overview: The circuit's noise channels (colored pink) are sampled to obtain the Error Realization (ER) for each channel and the combined ER of the circuit, which is an $n$-tuple of all individual channel ERs. If the ER $e_{i}$ occurs $s_{i}$ times, then the corresponding circuit $c_{i}$, is simulated once, and its output vector is sampled $s_{i}$ times. (B) ER Commutation overview: The ERs corresponding to the two samples are different - (X, II) vs (I, XX). However, they produce the same output statevector. Sample 1 can be converted into Sample 2 by commuting the X error through the CNOT (C) The circuits on the left and right are equivalent. However, we do not push the noisy X gate through the $R_{z}$ gate since it changes the argument from $\theta$ to $-\theta$, effectively changing the noiseless circuit. (D) We evaluate the significance of each circuit by comparing its frequency of occurrence $p_{i}$ with that of the most frequent circuit, $p_{max}$. If $p_i \geq \alpha\cdot p_{max}$, then it is significant, else it is not. All significant circuits are simulated. For insignificant circuits, we sample $\beta$ random circuits and add them to the set of circuits to simulate.}
    \label{fig:combined_ecm}
\end{figure*}

Another commonly used simulation paradigm is Tensor Network Simulation (TNS) \cite{orus2019tensor}. In spirit, it is similar to SVS since it also involves matrix vector multiplications. However, TNS defines an additional parameter, \textit{bond dimension} $D$, which is the maximum allowable rank of any matrix in the circuit. If the rank of a matrix $U$ is greater than $D$, we approximate it by assigning a value of 0 to its eigenvalues outside of the top $D$ eigenvalues. TNS enables efficient simulation of wider and deeper circuits compared to SVS. However, this comes at the cost of approximations in the output. TNS is especially useful for simulating low entanglement circuits since they naturally have a small value of $D$.

\section{Motivation and Proposal}\label{sec:motivation_and_proposal}

In Section \ref{sec:noisy_qcs}, we discuss that when using SVS or TNS, multiple simulations are needed to perform one noisy QCS. In effect, we trade off the memory overhead for higher computational cost. Our primary objective is to reduce this computational overhead and, hence, the simulation time. To achieve our objective, we introduce two modules - the \textit{Error Characterization Module} (ECM), and \textit{Depth First Tree Traversal} (DFTT). The ECM is further composed of three steps: (a) ER Tallying, (b) ER Commutation and (c) Pruning. The overall schematic of TUSQ is shown in Figure \ref{fig:tusq_overview}.

\ignore{
\begin{figure}[ht!]
    \centering
    \includegraphics[width=0.8\linewidth]{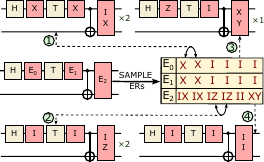}
    \caption{}
    \label{fig:er_tallying}
\end{figure}
}

\subsection{Error Characterization Module (ECM)}\label{subsec:ecm}
The goal of ECM is to analyze all circuit instances (say $S$) needed for a single noisy simulation and find instances that produce unique and significant outputs. Instead of computing the output statevector afresh every time, we compute only the unique instances and draw as many samples from them as their respective frequencies. Discarding insignificant circuits saves computational overhead at the cost of a marginal approximation error. To identify these instances, we propose three procedures: (a) Error Realization (ER) Tallying, (b) ER Commutation and (c) Pruning.

\subsubsection{\textbf{Error Realization (ER) Tallying}}\label{subsubsec:er_tallying}
As shown in Figure \ref{fig:qc_ideal_noisy_bg}(B), a circuit with noisy channels can be seen as a classical average of circuits with ``fixed noisy gates'' (the $I/X/Y/Z$ gates remain fixed). We can obtain these fixed circuits from the noisy circuits by sampling the noise channels. We call the set of gates sampled from the noise channels per SVS instance an \textit{error realization} (ER). For example, for a depolarizing noise channel with a $10\%$ error rate ($p=0.1$), the ER could be $I$ (No error) with $90\%$ probability ($1-p$), and $X$, $Y$, or $Z$ (Error) with probability $3.3\%$  each ($\frac{p}{3}$). For a circuit with $n$ depolarizing channels, the ER would be the $n$-tuple of the ERs of each channel, eg - $(I_{0}, X_{1}, Y_{2}, ..., I_{N-1})$. Currently, on most reasonable quantum computers, the error rate per operation varies between $0.1\%$ and $10\%$. This results in ERs with low Hamming weights \cite{hamming_weight} (fewer non-$I$ gates). Since the probability of the $I$ gate is much higher than the probability of $X/Y/Z$ gates, the lower the Hamming weight of an ER, the higher its probability of occurrence. Figure \ref{fig:qc_ideal_noisy_bg} (C) shows the frequency of ERs of length 20, as a function of their Hamming weights for $p=1\%$. We observe that the probability of an ER having a Hamming weight greater than 2 is zero.

Using this insight, we start tracking the frequencies of different ERs. We pre-sample all noisy channels for a given number of shots before performing the SVS, and keep track of the unique ERs and their frequencies. If any ER (say $e_{i}$) occurs $s_{i}$ times, we can simulate the corresponding circuit ($c_{i}$) once and sample the output state vector $s_{i}$ times rather than performing $s_{i}$ separate SVSs. Since sampling an output state vector multiple times is much cheaper than implementing matrix-vector multiplications \cite{patti2025augmenting}, ER tallying reduces compute overhead substantially. The high-level overview of ER tallying is shown in Figure \ref{fig:combined_ecm} (A).

\ignore{
\begin{figure}[ht!]
    \centering
    \includegraphics[width=0.8\linewidth]{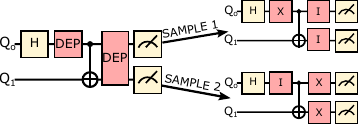}
    \caption{ER Commutation Overview: The ERs corresponding to the two samples look different - (X, II) vs (I, XX). However, they produce the same output statevector and are equivalent.}
    \label{fig:er_commutation}
\end{figure}
}

\ignore{
\begin{figure}[ht!]
    \centering
    \includegraphics[width=0.8\linewidth]{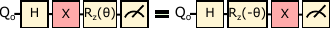}
    \caption{The circuits on the left and right are equivalent. However, we do not push the noisy X gate through the $R_{z}$ gate since it changes the argument from $\theta$ to $-\theta$, effectively changing the noiseless circuit}
    \label{fig:er_commutation_false_example}
\end{figure}
}

\subsubsection{\textbf{ER Commutation}}\label{subsubsec:er_commutation} ER tallying keeps count of the unique ERs and their frequencies. However, there are instances where, despite being different, the ERs produce the same output state vector. Consider the circuit shown in figure \ref{fig:combined_ecm} (B), and two sample ERs - $(X, II)$ and $(I, XX)$. Since these ERs are different, naively, we would compute the state vector for each of them separately. However, an X gate on the control qubit of a CNOT gate, when commuted through, leads to an extra X gate on the target qubit. This can be verified easily by performing the corresponding matrix multiplications. Thus, despite being different, these ERs produce the same result. Based on the circuit, this behavior might be ubiquitous, which gives us a greater opportunity for overhead reduction. ER commutation identifies such ER instances and groups them, i.e. if $(c_{1}, s_{1})$ and $(c_{2}, s_{2})$ are two (circuit, shot) tuples corresponding to distinct ERs $e_{1}$ and $e_{2}$, which produce the same output state vector, then we remove the $(c_{2}, s_{2})$ tuple and update $s_{1}$ as $s_{1} \rightarrow s_{1} + s_{2}$.

A noisy gate is pushed to the right until it encounters a situation where pushing it further will modify the noiseless circuit. For example, in Figure \ref{fig:combined_ecm} (B), we push the noisy $X$ gate through the CNOT since none of the noiseless circuit gates (Hadamard, CNOT, and measurement) is altered in this process. However, in Figure \ref{fig:combined_ecm} (C), we do not push the X gate through the $R_{z}$ since doing so would alter the argument of $R_{z}$ from $\theta$ to $-\theta$, modifying the noiseless circuit. Once this has been done, we simply compare the new ERs of $c_{1}$ and $c_{2}$. If they are the same, we add their counts.

\begin{figure*}
    \centering
    \includegraphics[width=0.95\linewidth]{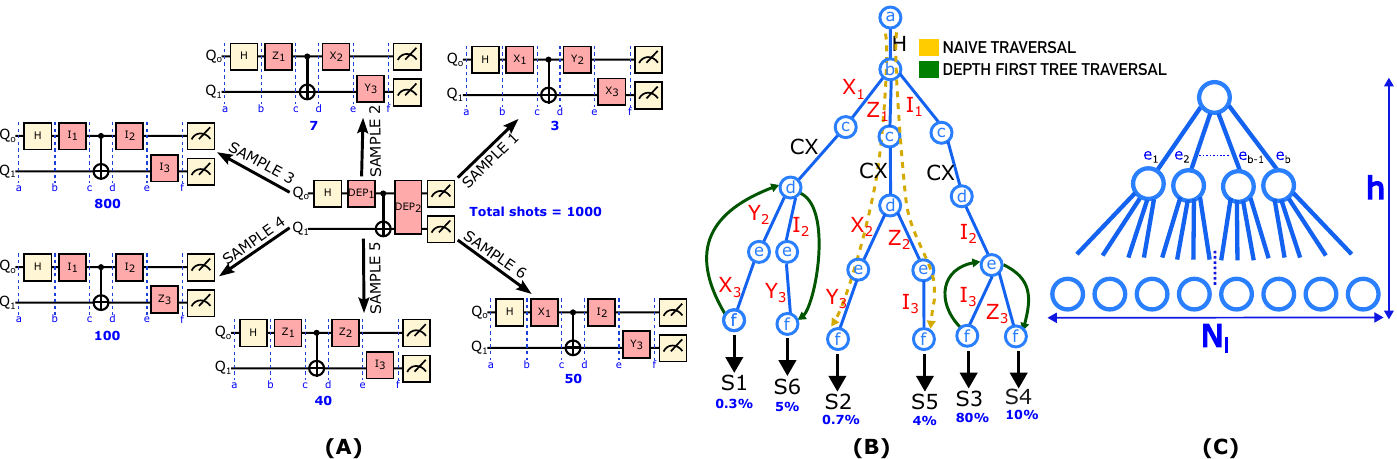}
    \caption{(A) Post-ECM circuits to simulate, along with their respective frequencies (B) The circuits are stored as a tree. The nodes represent the state vector at a particular point in the circuit. The edges represent gates. For each circuit, we can traverse from the root \textcircled{a} to the leaf \textcircled{f}, updating the state vector by multiplying it with the respective edge's gate (Naive traversal as shown by the golden arrows). Computation reuse is facilitated by rolling back to a previous intermediate state. For example, once we have computed the output of Sample 1 (S1), we traverse back the edges (apply inverse of the gate) to node (d), and then down the branch of S6 (green arrows). For the example shown using the green arrows, we don't have to perform the multiplications from node \textcircled{a} through node \textcircled{d} for S6. This is Depth First Tree Traversal (DFTT) (C) For a depolarizing noise model, every node has 4 child nodes. In the case of measurement noise, there are just 2 child nodes.}
    \label{fig:dftt}
\end{figure*}

\subsubsection{\textbf{ER Commutation - Algorithm and Implementation}}\label{subsubsec:er_comm_implementation}
In order to benefit from ER commutation, we must ensure that the time taken to push the gates is small. If each circuit consists of a total of $g_{1}$ gates out of which $g_{2}$ are noisy gates, then in the naive case we would need to perform $\mathcal{O}(g_{1} \cdot g_{2})$ operations per circuit to carry out ER commutation. The process would be similar to bubble sort. If the total number of unique circuits at the end of ER tallying is $S_{2}$, the total number of operations becomes $\mathcal{O}(S_{2}\cdot g_{1} \cdot g_{2})$. This is a high cost.

We can reduce this cost using a greedy algorithm that is very close to optimal in practice. The algorithm starts by initializing a vector of stacks, one stack corresponding to each qubit. Our circuits are assumed to be transpiled into a basis of single-qubit gates and the CNOT gate. The noise sources are measurement noise and depolarizing noise, which includes Pauli twirled decoherence noise (see Section \ref{subsec:compatible_noise}), all of which result in Pauli noisy gates. At every step of the algorithm, we maintain the invariant that each noisy gate is present as far to the right of the circuit as possible. This invariant is preserved using a set of commutation rules listed below:

\begin{enumerate}
    \item Multiple noisy Pauli gates back-to-back are multiplied together to condense them into a single noisy Pauli gate.
    \item A noisy Pauli gate can commute through any other noiseless Pauli gate.
    \item Pauli gates - $X/Y/Z$ commute through rotation gates along the respective axis, i.e. $R_{X}/R_{Y}/R_{Z}$.
    \item An $X$ gate on the control (target) qubit of a CNOT, when pushed through the CNOT, results in an $X$ gate on the control and target qubits (only on the target qubit) (see Figure \ref{fig:combined_ecm} (B)).
    \item A $Z$ gate on the target (control) qubit of a CNOT, when pushed through the CNOT, results in a $Z$ gate on the control and target qubits (only on the control qubit).
    \item A $Y$ gate on the control qubit of a CNOT, when pushed through the CNOT, results in a $Y$ gate on the control qubit and an $X$ gate on the target qubit, while a $Y$ gate on the target qubit of a CNOT, when pushed through, results in a $Z$ gate on the control qubit and a $Y$ gate on the target qubit.
\end{enumerate}

We iterate through the gates of our circuit. For each noiseless candidate gate in our circuit, we check the tops of the stacks of the qubits on which the gate acts. If the stack is empty or the top gate is one that doesn't commute with our candidate gate (as per our commutation rules), we push the candidate gate and proceed. If the top of the stack has a noisy gate that commutes (as per the commutation rules 2-6 enumerated below) with the candidate gate, then pop from the stack, push the candidate gate on the stack, and finally push new noisy gates according to the respective commutation rule. For each noisy candidate gate in the circuit, we check the tops of the stacks again. If we find existing noisy gates in the stacks, we simply merge the candidate gate with the existing noisy gates in the stacks according to rule 1; else, we push the candidate gate and proceed. This whole procedure maintains the circuit invariant of having noisy gates as much to the right as possible. 

This process is repeated until all gates in the circuit are exhausted. Once that is done, we combine the shots of the circuits with the same ER.

\subsubsection{\textbf{Pruning}}\label{subsubsec:pruning} Figure \ref{fig:qc_ideal_noisy_bg}(C) shows that the frequency of an ER decreases exponentially with an increase in its Hamming weight. Hence, the distribution of the number of ERs is not uniform. The primary contribution to an output distribution comes from a small fraction of ERs. We call circuits corresponding to these low hamming weight, high-frequency ERs as the \textit{significant circuits} ($\mathcal{C_S}$). The set of low-frequency circuits is called \textit{insignificant circuits} ($\mathcal{C_I}$). If the frequency of the most commonly occurring circuit is $p_{max}$, we define a constant $\alpha$ such that any circuit with frequency $p_{i} \geq \alpha \cdot p_{max}$ is significant, else it is insignificant. For this paper, we choose $\alpha=0.01$.

Since these circuits are supposed to signify noise, which itself has stochasticity associated with it, eliminating circuit instances with low weights would introduce only a negligibly small perturbation to the circuit. Hence, for the significant circuits, we obtain the output state vector and sample it normally. For the insignificant circuits, we represent their collective contribution, instead of their individual contribution (which is negligibly small). The individual frequency of each insignificant circuit may be small; however, the sum of their frequencies (say $p_{\text{insig}}$) can be substantial. For example, when simulating a 10-qubit QAOA circuit for a million shots, the significant circuits account for $58\%$, while the insignificant circuits account for $42\%$ of the total shots. In this case, simply ignoring all insignificant circuits introduces a large error. Hence, we randomly sample a subset of insignificant circuits $\mathcal{K}=\{(c_{t_{1}}, p_{t_{1}}), (c_{t_{2}}, p_{t_{2}}), ..., (c_{t_{\beta}}, p_{t_{\beta}})\}$. The number $\beta$ is a user-chosen hyperparameter. We compute the final state vector $v_{t_{i}}$ for each element in $\mathcal{K}$, and sample it for a total of $\frac{p_{insig}}{\sum_{i=1}^{\beta}p_{t_{i}}} \cdot p_{t_{i}}$ times. This ensures that the total contribution of the set of insignificant circuits is maintained in the output distribution, even though each circuit's individual impact on the output is not very noticeable. The final number of circuits that we need to simulate is given by 
\begin{equation}
    S_{final} = \sum_{i} \mathbf{1}_{p_{i} \geq \alpha\cdot p_{max}} + \min(\beta, |\mathcal{C_I}|)
\label{eq:s_final}
\end{equation}

\subsection{Depth First Tree Traversal (DFTT)}\label{subsec:dftt} Once we have obtained the set of circuits to simulate (the set of all significant circuits and representative insignificant circuits), we perform their SVS and average their output. The most naive way to achieve this is to perform matrix-vector multiplications for each SVS separately. However, we observe that there are many opportunities for computation reuse across circuits, which we exploit using DFTT.

Figure \ref{fig:dftt} (A) shows the post-ECM SVS candidate circuits along with their respective frequencies of occurrence. If we place these circuits in a tree-like structure as shown in Figure \ref{fig:dftt} (B), we can facilitate computational reuse across them. 

The nodes \textcircled{a}-\textcircled{f} denote the state vector at their respective positions in the circuit, while the edges represent gates acting on them (see Figure \ref{fig:dftt}(A)). A parent-to-child edge traversal involves updating the state vector by multiplying it by the gate corresponding to the edge between the nodes. A child-to-parent edge traversal involves multiplying the state vector by the inverse of the corresponding gate. For the time being, we assume that all gates, noisy or noiseless, are unitary and have valid inverses. Such cases are ubiquitous, as discussed in Section \ref{subsec:compatible_noise}. Circuits with non-unitary operations are handled in Section \ref{subsubsec:non_invertible_channels}. The output of a circuit can be computed by traversing the root (\textcircled{a}) to the leaf (\textcircled{f}) for that circuit.  

\begin{figure*}
    \centering
    \includegraphics[width=0.95\linewidth]{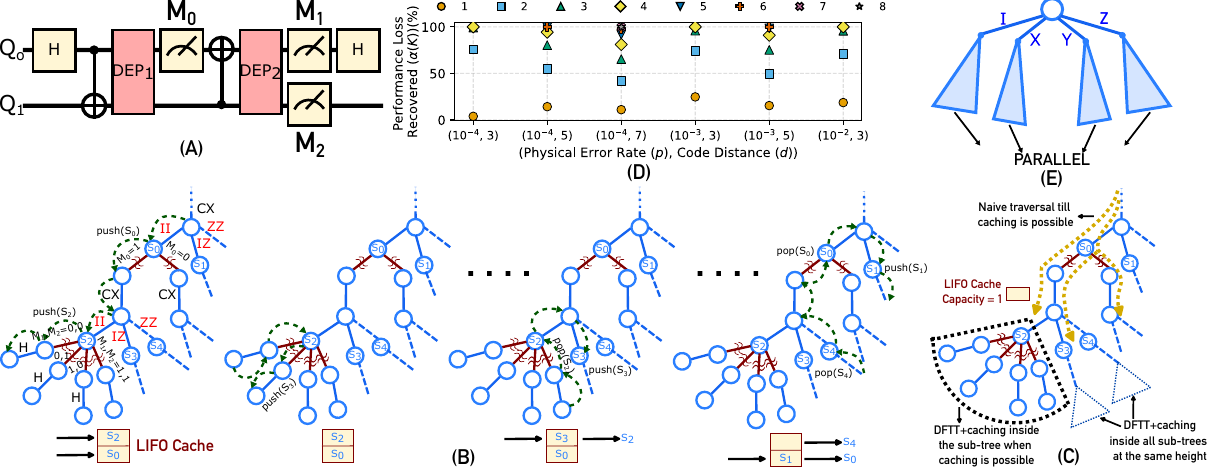}
    \caption{Refer to Section \ref{subsubsec:non_invertible_channels} for details on DFTT+Caching (A) Circuit with two layers of mid-circuit measurements that we simulate using DFTT+Caching. (B) Traversing the tree using DFTT+Caching for the circuit in subfigure (A). During tree traversal, pre-mid-circuit-measurement (MCM) nodes are pushed to a LIFO cache (capacity $K$). These are accessed again when we traverse back through the mid-circuit measurement edge (colored red and marked with the $\approx$ symbol). We annotate the edges of the first tree with the names of the operation they correspond to. We do not draw the complete tree due to space constraints and represent potentially omitted edges using dashed blue lines. Although demonstrated for mid-circuit measurements, DFTT+Caching works for all non-invertible channels. (C) When cache capacity (say $K=1$) is less than the number of MCM layers (two layers for our circuit), we traverse down to the $K$-th pre-MCM node from the leaf (shown with yellow dotted arrows) and perform DFTT+Caching on the sub-tree rooted at this node. We do this for all eligible subtrees. (D) Non-invertible channels force us to ``switch-off'' DFTT and use naive tree traversal, causing performance loss. Caching states before non-invertible channels and using DFTT for invertible channels enables us to recover performance (see Section \ref{subsubsec:non_invertible_channels} for details). We show the performance recovered as a function of the number of caches for six different surface code circuits for physical error rate $(p)=10^{-2}, 10^{-3} \text{ and } 10^{-4}$ and code distances $(d) =3,5,7$. (E) If extra memory is available, we parallelize DFTT across sub-trees. An $n-$fold parallelization leads to an $n-$fold increase in memory.}
    \label{fig:dftt_with_caching_extended}
\end{figure*}

Once we have reached a leaf, we can reuse parts of the computation to reach other leaves.  For example, in Figure \ref{fig:dftt}(B), once we have reached the output of S1, we can follow the green arrows to obtain the output of S6, without having to traverse nodes \textcircled{a}-\textcircled{d} again. We roll back to a previous state and proceed from there instead of starting from scratch for S6, which saves us many compute operations. As the circuit gets larger, these benefits multiply. As mentioned in Section \ref{sec:introduction}, an alternative approach, as taken by TQSim \cite{wang2025accelerating}, \cite{li2020eliminating}, would be to memoize node \textcircled{d}  when computing S1, and use it later when computing S6 instead of starting from scratch. Ref \cite{wang2025accelerating} also discusses how the memory in large-scale HPC systems stays underutilised to motivate their memoization based solution. In the case of DFTT, this additional available space is utilized by traversing multiple sub-trees in parallel. For example, if we are utilizing only $25\%$ of the total memory, we can copy the state in the root down to its children and perform DFTT on the respective subtrees in parallel, as shown in Figure \ref{fig:dftt_with_caching_extended} (E). This offers an additional speedup.

\subsubsection{\textbf{DFTT-Asymptotic Complexity}}\label{subsec:dftt_complexity}

In this section, we formally show the asymptotic advantage obtained by performing DFTT instead of naive simulation.

Consider a noise model where after every gate we apply a noisy channel with $b$ different possibilities $\{e_{1}, e_{2}, ..., e_{b}\}$. For example, in the case of single qubit depolarizing noise b = 4 ($\{I, X, Y, Z\}$), and in the case of measurement noise b = 2 ($\{I, X\}$). For simplicity, we will assume the value of $b$ to be fixed throughout the circuit. This creates a tree in which the number of nodes at each tree level grows exponentially. Let the total number of edges in the tree be $|E|$, the height of the tree be $h$, and the number of leaf nodes be $N_{l}$ (see Figure \ref{fig:dftt} (C)). For DFTT, the number of operations ($T_{dftt}$) is equal to the number of edge traversals. We traverse every edge twice, which gives us $T_{dftt}=2|E| = \mathcal{O}(|E|)$. In the naive implementation, we traverse all the way from the root to a leaf node for every leaf node. The number of such traversals is equal to the number of leaves -  $N_{l}$. The number of edges in each traversal is equal to the height of the tree. Hence $T_{naive} = N_{l} \times h$. Note that, at every tree level, the number of edges grows exponentially by a factor of $b$. Hence, $b + b^{2} + ... + b^{h} = |E| \implies b\cdot(\frac{b^{h} - 1}{b - 1}) = |E|$\\
$\implies h = \log_{b}((b-1)|E|+b)-1$\\
Further, the number of leaf nodes is equal to the number of edges at the last level, which gives us $N_{l} = b^{h} = (1 - \frac{1}{b})|E|+1$. 

Using these two expressions, we get $T_{naive} = N_{l} \times h = ((1 - \frac{1}{b})|E|+1) \times (\log_{b}((b-1)|E|+b)-1) = \mathcal{O}(|E|\log_{b}|E|)$. 
Thus, DFTT reduces the number of operations from $\mathcal{O}(|E|\log_{b}|E|)$ to $\mathcal{O}(|E|)$.

\subsubsection{\textbf{Simulation of non-invertible channels}}\label{subsubsec:non_invertible_channels}

While the ECM works well for arbitrary quantum channels, the backtracking step in DFTT enforces all operations to be invertible. This assumption, although mostly valid, breaks down in certain key scenarios, e.g., mid-circuit measurements and erasures. A naive approach to simulating such circuits within TUSQ is to ``switch off'' DFTT, where we perform a root-to-leaf traversal for each leaf node independently, as shown by the golden dotted arrows in Figure \ref{fig:dftt} (B). Hence, the simulation doesn't fail, but the speedup from DFTT disappears. All speedup is due to the ECM alone. 

However, based on the amount of memory available, we can tweak the design of DFTT to incorporate caching, thereby offering a speedup even in the presence of non-invertible channels. If we cache the state immediately before a non-invertible channel, DFTT can proceed normally throughout the circuit tree, except when backtracking across a non-invertible edge. Instead of backtracking, which is not permitted in this case, we fetch the state from the cache. We call the set of states to be cached the ``caching set''. We explain our design, which we call \textit{DFTT+Caching}, with the help of Fig. \ref{fig:dftt_with_caching_extended}. Fig. \ref{fig:dftt_with_caching_extended}(A) shows the circuit we are simulating. It consists of two layers of mid-circuit measurements (MCMs). Note that while there are three MCMs in the circuit, the number of non-invertible channels is just two - one corresponding to each layer of MCMs. We can merge all MCMs in the same layer into a single edge in the tree since access to a state with only a subset of concurrent MCMs applied (and others not applied) is never needed in our simulation. Merging all concurrent MCMs reduces the cache requirements drastically. Figure \ref{fig:dftt_with_caching_extended} (B) shows the DFTT+Caching protocol for a cache size, $K=2$. All the edges corresponding to non-invertible MCM layers are colored red and marked with the $\approx$ sign. As soon as we encounter a pre-MCM node in the tree, we push it into a Last-In-First-Out (LIFO) cache. We pop the node once we have back-traversed all of its child edges (i.e. reached the node back from all of its children) since it would not be needed subsequently in the simulation. We perform the tree traversal while pushing to and popping from the LIFO cache till we reach the rightmost leaf of the tree, which marks the end of the simulation.

Note that in Fig. \ref{fig:dftt_with_caching_extended}(B), the cache capacity, $K=2$, is equal to the number of non-invertible channels in the circuit. Hence, all pre-MCM nodes make it to the caching set. This enables us to fully recover any potential performance loss caused by non-invertible channels. However, in the memory-constrained regime, caching opportunities are limited. Often, the cache capacity, $K$, is smaller than the number of non-invertible channels. In this case, the caching set stores the $K$ pre-MCM nodes that are closest to the leaf in every branch. Figure \ref{fig:dftt_with_caching_extended} (C) shows this situation with cache capacity $K=1$. Here, in the first branch, state $S_2$ gets cached while $S_0$ doesn't (since $S_2$ is closer to the leaf than $S_0$). Since we cannot perform DFTT+Caching throughout the tree in this case, we traverse down to the ``shallowest'' (one with the least depth or farthest away from a leaf) node in the caching set from the root, and perform DFTT+Caching on the sub-tree rooted at this node. In Figure \ref{fig:dftt_with_caching_extended} (C), the ``shallowest nodes'' are $S_2, S_3$ and $S_4$. Once we finish DFTT+Caching inside one sub-tree, we traverse down to another ``shallowest node'' from the root and traverse its sub-tree using DFTT+Caching. This process is repeated till all sub-trees are traversed.

Figure \ref{fig:dftt_with_caching_extended}(D) shows the performance recovered by DFTT+Caching ($\alpha(K)$) as a function of cache capacity ($K$) for rotated Surface Code memory circuits. We use the circuits' inbuilt Stim implementation \cite{gidney2021stim}. We sweep the physical error rate ($p$) across three values - $10^{-2}, 10^{-3}, 10^{-4}$ and the code distance ($d$) across three values too - $3, 5, 7$, leading to circuits with $26/64/118$ physical qubits. We perform $d$ rounds of measurement on each circuit, which leads to $d$ non-invertible channels. Each circuit is sampled 1 million times. Suppose that the traversal with ``DFTT switched off'' requires $N_1$ operations. Now, consider a hypothetical scenario where a tree with non-invertible edges can be fully traversed using DFTT in $N_2$ operations. Note that this scenario is not realistic; however, $N_2$ serves as a tight lower bound for DFTT+Caching. The number of operations performed by DFTT+Caching for a $K$-sized cache is denoted by $N_{DFTT+Caching, K}$. Our performance recovery metric $\alpha(K)$ is defined as
 
\begin{equation}
    \alpha(K) = \frac{N_1 - N_{DFTT+Caching, K}}{N_1 - N_2}
\end{equation}

When $N_{DFTT+Caching, K}$ is equal to the $N_2$ (operations needed by DFTT), we recover $100\%$ performance.

Note that for Figure \ref{fig:dftt_with_caching_extended}(D), we do not perform the actual traversal. Rather, we count the number of operations needed to traverse the tree, enabling us to get estimates for circuits with up to 118 qubits. A single-qubit matrix vector multiplication is counted as 1 operation, and a two-qubit matrix vector multiplication is counted as 4 operations. Forward traversal across a non-invertible edge is 1 operation, while a back traversal is 0 operations (since we fetch from the cache).

We observe that a capacity-3 cache is enough to recover $60\%-$  $100\%$ of the performance across benchmarks. To recover the same amount of performance, the cache requirement increases with $d$. A higher $d$ implies more measurement rounds, and hence more non-invertible channels, which leads to a larger cache requirement. 

Other non-invertible channels like erasures, leakage errors, etc., can also be simulated using DFTT+Caching. Such logical channels are ubiquitous in FTQC-relevant logical level simulations, further highlighting the usefulness of TUSQ.

\section{Methodology}\label{sec:methods}

\begin{figure*}[h!]
    \centering
    \includegraphics[width=0.96\linewidth]{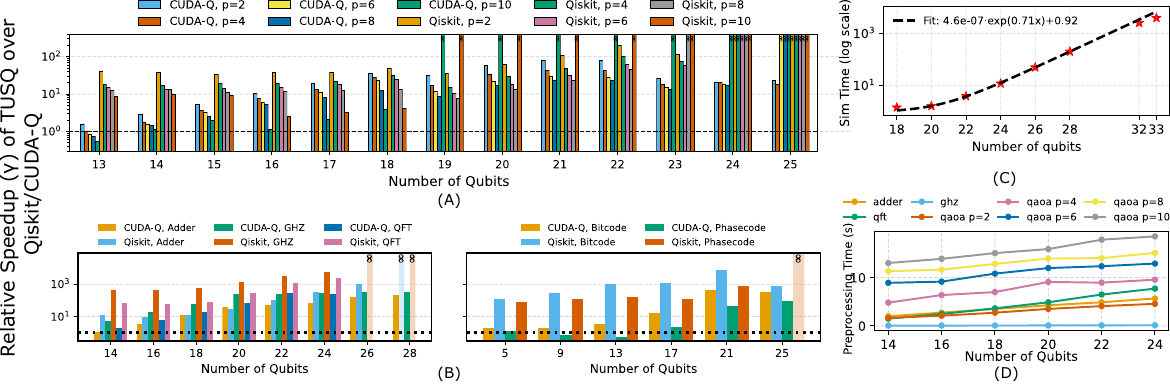}
\caption{
Speedup Offered by TUSQ over Qiskit and CUDA-Q for QAOA (Subplot A), and Adder, GHZ, QFT, Bitcode and Phasecode (Subplot B). Note that for the QAOA subplot, $p$ represents the number of layers in the QAOA ansatz. Higher $p$ implies a deeper circuit. The bar's height represents the time taken by the baseline divided by TUSQ's wall time. The higher the bar, the greater the speedup. There are cases where the baseline times out on the Perlmutter cluster. These are represented with a bar reaching the roof of the plot and marked with an $\infty$ sign. TUSQ is the fastest simulator for 193 out of 198 benchmarks. There are 5 cases where CUDA-Q outperforms TUSQ (9, 13 qubit Phasecode and 13 qubit,p=6/8/10 QAOA). These programs have a small simulation time and are not time-critical. In this case, TUSQ's CPU pre-processing time becomes the speed bottleneck instead of GPU execution time. However, the wall clock time is low in both cases, with TUSQ simulating these programs in 1.36, 4.49, 29.37, 44.20, and 77.14 seconds, respectively. Compared to Qiskit, TUSQ's average relative speedup is $59.06\times$, with a maximum value of $7878.03 \times$. Compared to CUDA-Q, the relative speedup is $13.38\times$ with a maximum value of $439.38 \times$.
(C) Time taken to simulate 32 and 33 QAOA circuits (p=2, 32k shots) by splitting the statevector across multiple GPUs. Smaller ($< 30$ qubits) circuits need only a single GPU. The simulation time for 32 and 33-qubit circuits fits the exponential curve generated from the smaller circuits, demonstrating that multi-GPU execution does not add any significant serial overhead in TUSQ. (D) CPU preprocessing time of TUSQ: We observe an expected increase in the CPU pre-processing time with qubit size, with an average and maximum time of $3.97$ and $18.52$ seconds, respectively.}
    \label{fig:all_tusq_evaluations}
\end{figure*}

\subsection{Evaluation Infrastructure}\label{subsec:infrastructure}
TUSQ is evaluated for noisy Statevector and Tensor Network simulations. TUSQ uses Nvidia's cuStateVec and cuTensorNet libraries \cite{bayraktar2023cuquantum} as the backend kernels for performing statevector and tensor network simulations, respectively. Note that since TUSQ is backend agnostic, the user can use any simulation kernel of their choice.  
All benchmarking experiments were performed on the NERSC \textit{Perlmutter} supercomputer. Each compute node is equipped with an AMD EPYC 7763 CPU (64 cores/128 threads), and NVIDIA A100 (40 GB) GPUs. Unless explicitly stated, computations were restricted to a single GPU using the \texttt{CUDA\_VISIBLE\_DEVICES=0} environment variable.

\subsection{Benchmarks}\label{subsec:benchmarks}
For our statevector simulations, we use the benchmarks from the Supermarq \cite{Supermarq} benchmark suite given in Table \ref{tab:benchmark_details}. Our benchmarks have been chosen to represent a variety of circuit structures - linear (GHZ, Bitcode, Phasecode), and parallel (QAOA); qubit counts - 13-28 qubits; circuit depths - 4-770; gate counts - 4-1250; and output distributions - unimodal (Adder, Bitcode, Phasecode), bimodal (GHZ), distribution with pronounced peaks (QAOA), and uniform distribution (QFT). 

For the multi-GPU evaluation study (Section \ref{subsec:simulator_speed} and Fig. \ref{fig:all_tusq_evaluations} (C)), we simulate up to 33-qubit circuits. For our tensor network simulation study (Section \ref{subsec:tusq_utility} and Table \ref{tab:large_qubit_estimate}), we simulate up to 40-qubit circuits.

\begin{table}[]
    \centering
    \begin{tabular}{|c|c|c|c|}
    \hline
         Benchmark & Qubits & Depth & Gate Counts \\
         \hline
         QAOA & 13-25 & 82-770 & 130-1250 \\
         \hline
         Adder & 4-28 & 69-289 & 97-417 \\
         \hline 
         Bitcode & 5-25 & 4-144 & 4-144 \\
         \hline
         Phasecode & 5-25 & 8-48 & 20-470 \\
         \hline
         GHZ & 14-28 & 14-28 & 14-28 \\
         \hline 
         QFT & 14-24 & 27-47 & 105-300 \\
         \hline
         BV & 4-24 & 6-26 & 14-74 \\
         \hline
         
    \end{tabular}
    \vspace{6pt}
    \caption{Benchmarks used to evaluate TUSQ for noisy statevector simulations}
    \label{tab:benchmark_details}
\end{table}

\subsection{Metrics}\label{subsec:metrics}
\subsubsection{Speedup ($\gamma_{A/B}$)} This metric quantifies the simulation speedup of protocol ``A'' with respect to protocol ``B'' as the ratio of the time taken by protocol ``B'' to that of the time taken by protocol ``A''. 

\begin{equation}
    \gamma_{A/B} = \frac{simulation\_time_{B}}{simulation\_time_{A}}
\end{equation}

\subsubsection{Relative Fidelity Difference ($\delta_{A, B}$)} Relative Fidelity Difference quantifies the deviation in fidelity reported by protocols $A$ and $B$ for the same program. If the fidelity of protocol $A$ is $f_{A}$, and fidelity of protocol $B$ is $f_{B}$, we have 
\begin{equation}
    \delta_{A, B} = \frac{|f_{A} - f_{B}|}{f_{A} + f_{B}}
\end{equation}

where $0<f_{A}, f_{B}\leq 1$ and $0 \leq \delta_{A, B} \leq 1$.

\subsection{Noise Modeling}\label{subsec:noise_modeling}
Our current noise models involve depolarising errors, measurement errors, and amplitude and phase damping errors, with the latter two being represented in their Pauli twirled form \cite{ghosh2012surface}. The Pauli twirling is needed to ensure Unitary noise gates, which are a prerequisite for DFTT. A detailed discussion of compatible noise models and their practical usefulness is given in Section \ref{subsec:compatible_noise}. For our experiments, we consider a value of probability of error $p=1\%$. We also evaluate TUSQ for $p=0.1\%$ for some studies. 

\subsection{Baseline}\label{subsec:baseline}
We compare TUSQ against three GPU noisy simulators -  Nvidia's \texttt{CUDA-Q} version 0.11.0, IBM's \texttt{Qiskit} version 2.1.0 and TQSim \cite{wang2025accelerating}. These first two simulators are widely adopted in the community, while TQSim is a recently published work. The backend kernel in all libraries is \texttt{Nvidia cuStateVec v1.12.0}. For our Tensor Network Simulation study, we integrate TUSQ with \texttt{Nvidia cuTensorNet v2.9.1} and compare it against CUDA-Q with the \texttt{tensornet-mps} flag.

\subsection{Number of Shots}\label{subsec:num_shots}
There are various factors which decide the number of shots needed for simulation. We need to choose enough samples such that the statistical error with respect to DMS is minimized. Another factor is the output circuit fidelity. Since we are performing a noisy simulation, we want enough shots such that our output distribution has some useful signal. Our largest circuit has 1250 gates. For a $1\%$ error rate, the fidelity is $3.4\times10^{-6}$. Hence, we need at least $10^6$ shots to get a useful signal. For a lower error rate, we need fewer shots to get a useful signal. Another important factor to consider is the sampling error when measuring the quantum state. We want to have enough shots such that the quantum state is accurately captured. However, the exact number of shots here depends on the shape of the output distribution as well. In the case of workloads like BV, which have a single peak, the number of shots needed to capture the quantum state is low. However, for QFT or QAOA, where information about the entire distribution is needed, we need a large number of shots.

Past work also seems to have no consensus on the number of shots. Wang et. al. \cite{wang2025accelerating} use 32,000 shots for all workloads (up to 25 qubits), while Patti et. al. \cite{patti2025augmenting} use $10^{9}$ shots for a 35 qubit simulation. For our work, we show our results for a variety of shots - $32$k, $100$k, $1$ Mil, and $10$ Mil.

\section{Evaluation}\label{sec:evaluation}

\subsection{Speedup}\label{subsec:simulator_speed}
Figure \ref{fig:all_tusq_evaluations} (A)-(B) shows the performance of TUSQ compared to Qiskit and CUDA-Q for 198 benchmarks. The height of each bar equals the value of the relative speedup. Bars that touch the roof of the plot and are marked with an $\infty$ sign correspond to cases where the baseline (Qiskit/CUDA-Q) didn't finish in 40 hours and timed out on Perlmutter.

TUSQ outperforms Qiskit on all benchmarks, and CUDA-Q on 193 out of 198 benchmarks. TUSQ is worse than CUDA-Q for the 9 and 13 qubit Phase Code, and the 13 qubit QAOA with p=6, 8, and 10. These programs have small simulation times and are not time-critical. In this case, TUSQ's CPU pre-processing time becomes the speed bottleneck instead of GPU execution time, hence a conventional GPU simulation is faster. However, the wall clock time is low in both cases, with TUSQ simulating these programs in 1.36, 4.49, 29.37, 44.20, and 77.14 seconds, respectively.

We observe that TUSQ's relative speedup increases with the number of qubits. This is expected since simulation time scales exponentially in the number of qubits. Hence, any improvement in simulation time also scales exponentially.  Another trend that we observe in Figure \ref{fig:all_tusq_evaluations} (A) is that keeping the number of qubits fixed, the speedup reduces with an increase in the value of the depth parameter $p$ (not to be confused with simulation error rate -- $p$ is an overloaded term in literature). This happens because of the reduction in the effectiveness of ER Tallying at greater depths. A deeper circuit results in a higher likelihood of ERs being distinct for the same number of qubits. This is also reflected in Figure \ref{fig:combined_ablation_shot_variation} (A), where the effectiveness of ER Tallying drops dramatically between the $p=2$ and $p=6$ case for all QAOA workloads.  

Compared to Qiskit, TUSQ's average relative speedup is $59.06\times$, with a maximum value of $7878.03 \times$. Compared to CUDA-Q, the relative speedup is $13.38\times$ on average with a maximum value of $439.38 \times$.
Figure \ref{fig:all_tusq_evaluations}(D) shows the scaling of CPU pre-processing time for TUSQ. It increases with the number of qubits, with an average and maximum time of 3.97 and 18.52 seconds, respectively. TUSQ can be extended to multi-GPU and multi-node systems for large statevectors. We split the statevector into multiple sub-vectors that are stored across different GPUs. If a gate acts only on one sub-vector, we update it locally. If it involves the manipulation of amplitudes scattered across multiple sub-vectors, we move all relevant amplitudes to a single GPU and operate on them. This incurs a communication overhead on top of the compute cost. Figure \ref{fig:all_tusq_evaluations}(C) shows the time taken by TUSQ for 32 and 33-qubit simulations across multiple GPUs. We observe that these simulation times fit the exponential curve generated by the simulation times of smaller ($<30$ qubit) circuits, demonstrating that multi-GPU execution does not add any significant serial overheads.

\subsection{Deviation in Fidelity}\label{subsec:fidelity_deviation}

\begin{figure}
    \centering
    \includegraphics[width=\linewidth]{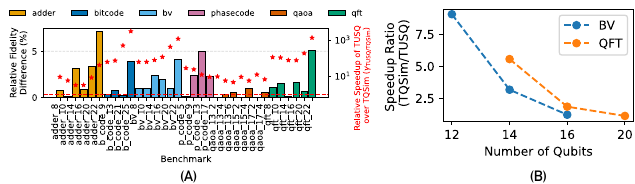}
    \caption{(A) Primary axis - Relative fidelity difference due to Pruning. Pruning is the only potential source of fidelity loss in TUSQ. All other steps are perfectly fidelity preserving. We see a relative fidelity difference of $1.66\%$ on average, which goes up to $7.15\%$. Secondary axis (in red) - Relative speedup of TUSQ over TQSim (in the high compute and memory critical regime), keeping fidelity loss the same for both methods. The fidelity loss incurred is equal to the bar height. TUSQ consistently outperforms TQSim for the same value of fidelity loss with an average and maximum spedup of $39.32\times$ and $3134.31\times$ respectively (B) Relative speedup of TQSim over TUSQ in the low compute, non memory-critical regime. For BV and QFT, TQSim is on average $3.26\times$ and $2.25\times$ faster than TUSQ, respectively. As the number of qubits grows (and the benchmarks become memory-critical), the speedup ratio decreases, demonstrating that TUSQ is the optimal choice for time and memory-critical benchmarks.}
    \label{fig:comparison_with_tqsim}
\end{figure}

\begin{figure}[ht!]
    \centering
    \includegraphics[width=0.85\linewidth]{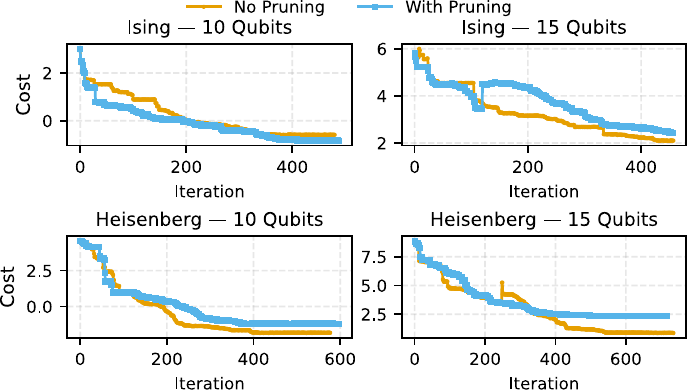}
    \caption{VQE convergence plots for 10 and 15 qubit Ising and Heisenberg Hamiltonians, with and without pruning error. The plots in all cases show similar convergence, highlighting the negligible effect of pruning error on VQE algorithmic correctness.}
    \label{fig:vqe_pruning_error}
\end{figure}

The Pruning step in TUSQ eliminates ``insignificant'' circuits which might lead to a deviation in fidelity from the baseline. All other steps of TUSQ - ER Tallying, ER Commutation, and DFTT are fidelity preserving. Figure \ref{fig:comparison_with_tqsim} (A) (primary axis) shows the Relative Fidelity Difference between the Pruning and no-Pruning approaches ($\delta_{pruning, no\_pruning}$) for six benchmarks - BV, Adder, Bitcode, Phasecode, QAOA, and QFT. These values have been calculated for $\alpha=0.01$ and $\beta=100$. The size of benchmarks varies from 8 to 22 qubits. The average value (arithmetic mean) of $\delta$ is $1.66\%$, and the maximum value is $7.15\%$. Note that geometric mean is not a valid metric in this case, since $\delta$ is equal to 0 for a few cases. 

The effect of these fidelity deviations on algorithmic correctness is minimal, if any. For example, as shown in Figure \ref{fig:vqe_pruning_error}, the VQE convergence plots with and without pruning error for 10 and 15-qubit Ising and Heisenberg Hamiltonians \cite{ising, heisenberg} are very similar. For the Adder and BV benchmarks, the final output value from a noisy distribution is the bit-string with the highest frequency. We compare 320 instances of the Adder circuit (corresponding to 4-22 qubits) with and without pruning error. We observe that for 289 out of 320 instances, the inferred bit-string from a pruned distribution is the same as the one inferred from the original unpruned distribution. We also compare 380 instances of the BV benchmark (with and without pruning error), for qubits ranging from 4 to 26. We observe that the output bit-string is the same for the pruned and unpruned distributions in 368 out of 380 cases. Hence, we see that for most cases, the pruning error does not have a practical effect on algorithmic correctness.

Although a strict trend for $\delta$ cannot be predicted because of its dependence on many variables, we generally expect its value to increase as the number of gates in the circuit increases. This is because more gates imply a greater number of error channels, which makes the ER frequency distribution less skewed. A less-skewed probability distribution deviates more from the original when we chop off its tail.

Another thing to note is that lower $\delta$ values can be achieved at the cost of increased simulation time. If the user wants less fidelity deviation, then a lower value of $\alpha$ and a higher value of $\beta$ parameters should be used. The exact values to be used are based on user preferences.

\subsection{Scaling with shots and physical error rate}\label{subsec:shots_scaling}

Figure \ref{fig:combined_ablation_shot_variation} (B) shows the speedup $(\gamma_{TUSQ/CUDA-Q})$ of TUSQ relative to CUDA-Q for shots ranging from 32,000 to 10,000,000. As mentioned in Section \ref{subsec:num_shots}, the number of shots needed to run a workload depends on various factors. To demonstrate the effectiveness of TUSQ across shot values, we perform a sweep from 32,000 to 10,000,000 shots. 

We observe that $\gamma_{TUSQ/CUDA-Q}$ increases with the number of shots. An increase in shots comes with added compute time. However, it also means that the likelihood of any two circuits having computational redundancies increases. Since CUDA-Q doesn't look for opportunities to eliminate redundancies, it encounters a sharper increase in compute time relative to TUSQ. Hence, we report a greater speedup with more shots, further highlighting the performance of TUSQ for compute-intensive, time-critical benchmarks. We would like to highlight that while the performance gains of TUSQ increase with the number of shots, it remains the better choice across the entire range of shots considered. The speedup numbers for $p=1\%$ (solid markers) are given in Table \ref{tab:speedup_variation_with_shots}. 

For $32k$ and $100k$ shots, we also evaluate the speedup for $p=0.1\%$ (hollow markers). For all benchmarks, the speedup in the case of $p=0.1\%$ is greater than in the case of $p=1\%$. This is attributed to a higher likelihood of the $I$ gate in ERs. This would mean more computational redundancy and hence, a higher speedup. 

\begin{table}[t]
\centering
\resizebox{0.8\columnwidth}{!}{
\begin{tabular}{|c|c|c|c|c|}
\hline
Shots & 32k & 100k & 1 Mil & 10 Mil \\
\hline
$\gamma_{TUSQ/CUDA-Q}$ 
  & $4.51\times$ 
  & $14.1\times$ 
  & $140.51\times$ 
  & $2075.44\times$ \\
\hline
\end{tabular}
}
 \vspace{6pt}
\caption{Variation of $\gamma_{TUSQ/CUDA-Q}$ for different shot values.}
\label{tab:speedup_variation_with_shots}
\end{table}

\subsection{Ablation study}\label{subsec:ablation_study}

\begin{figure}
    \centering
    \includegraphics[width=0.93\linewidth]{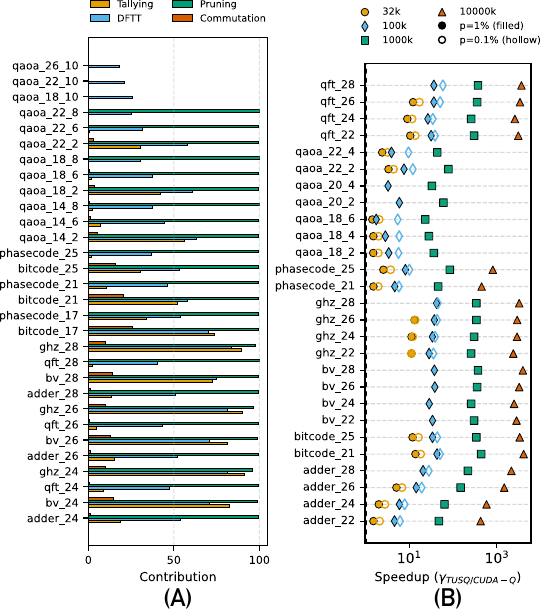}
    \caption{(A) Cost Reduction by different components of TUSQ across various benchmarks. (B) Variation of TUSQ's speedup over CUDA-Q, for shots ranging from 32K to 10 Million, for $p=1\%$ and $p=0.1\%$.}
    \label{fig:combined_ablation_shot_variation}
\end{figure}

Figure \ref{fig:combined_ablation_shot_variation} (A) shows the relative speedup provided by each optimization of the TUSQ pipeline. The speedup is quantized by the percentage reduction in matrix-vector multiplications achieved by each stage in isolation. Note that the percentage reduction achieved by all four stages in conjunction with each other is less than the sum of the reduction achieved by the four stages individually. This is because different optimizations may end up eliminating the same source of redundancy in the simulation, i.e. they have overlapping regions of optimization.

We see that DFTT consistently achieves a substantial speedup across all benchmarks corresponding to the elimination of the $\mathcal{O}(\log|E|)$ factor. For short to medium depth benchmarks, ER Tallying's speedup is large. It reduces as the depth increases. This is expected since more gates result in a longer ER, which reduces the chances of repeated samples that can be batched together. We see ER Tallying produce speedups at par with Pruning. Most importantly, unlike Pruning, they don't cause fidelity loss. ER Commutation provides the least amount of speedup, which also decreases with circuit depth for the same reasons as ER Tallying.

The speedup provided by Pruning is large as long as the output distribution is ``peaked'' - i.e. we can distinguish signal from noise. However, for very deep circuits - e.g. $QAOA, p=10$, where the signal is totally washed out, we observe that pruning is totally ineffective. These circuits are so deep that almost every shot produces a unique ER, eliminating any distinction between ``significant'' and ``insignificant'' ERs. Unlike the other three optimizations, the speedup from Pruning comes at the cost of deviation in fidelity as discussed in Section \ref{subsec:fidelity_deviation}. Note that the numbers reported in Figure \ref{fig:combined_ablation_shot_variation}(A) are for $\alpha = 0.01$ and $\beta = 100$. Since $\alpha$ and $\beta$ are user-defined parameters dependent on the user's error tolerance, Pruning's speedup becomes a function of this error tolerance.

Across all evaluated benchmarks, the average (maximum) speedup of each component w.r.t. the previous ones is $30.51\% (91.19\%)$ for ER Tallying, $5.14\% (25.92\%)$ for ER Commutation, $89.32\% (99.79\%)$ for Pruning, and $50.79\% (83.58\%)$ for DFTT. On average, the speedup from lossless optimizations (ER Tallying, ER Commutation and DFTT) is $46.62\%$, while that from Pruning is $53.3\%$. Note that we report the arithmetic mean here because some entries in our data are 0, which makes GM an infeasible metric.

\subsection{Comparison Against TQSim}\label{subsec:tqsim_comparison}

TQSim, a recent noisy simulator by Wang et. al. \cite{wang2025accelerating} also uses a tree to speed up noisy simulations. Despite peripheral similarities, there are marked differences in the design and performance of the two methods. TQSim performs simulation using a combination of breadth-first search, memoization and sampling. It starts by chopping a $n$ qubit, depth $d$ circuit ($C(n,d)$) into $k$ disjoint sub-circuits $\{sc_{i}(n,d_{i}), i \in [0,1,...k-1]\}$ such that each sub-circuit has $n$ qubits and the sum of depths of all sub-circuits is $d$, i.e. $\sum d_{i} = d$. The tree's root node represents the start of the circuit. Nodes at depth $i$ correspond to the ensemble of statevectors at the end of subcircuit $i$ that represents all possible ERs. TQSim samples a subset of representative nodes at different tree depths and memoizes their statevector until memory saturation. The saved states can be reused for future computation, which reduces simulation time. Whenever the memory saturates before all intermediate states can be saved, there is an inevitable loss in simulation fidelity. Since memory saturation is a common phenomenon, especially for large statevectors, TQSim proposes methods to cache ``important'' nodes that lead to minimal fidelity loss while also maximising simulation speed.   

In contrast, TUSQ traverses the tree in a depth-first fashion. No memoization is needed (unless simulating non-invertible channels), making our speedup purely algorithmic, unlike TQSim, where the speedup is a function of available memory. TUSQ uses ERs as an intermediate representation (IR) to determine whether two circuits give the same output without actually computing it. The ERs let us spot redundant computation (which can be eliminated without any fidelity loss), and insignificant computation (which can be eliminated at a small fidelity loss - the aggressiveness being a tunable knob in the form of $\alpha$ and $\beta$ parameters). This is in contrast to TQSim, where no IR is used. Any reduction in computation is accompanied by an inevitable loss in fidelity. The sweet spot between fidelity loss and speedup is determined using statistical methods.  

To evaluate the performance of TUSQ against TQSim in the time and memory critical regime, we compute the relative speedup ($\gamma_{TUSQ/TQSim}$) for six benchmarks - adder, bitcode, bv, phasecode, qaoa, and qft. The number of qubits ranges from 5 to 25. For our evaluation, we use an error rate $p=1\%$, 1 million shots, and a GPU memory of 40 GB for all experiments. The results are shown in Figure \ref{fig:comparison_with_tqsim} (A) on the red secondary axis. For both simulators, we use the same error budget (given by the bar plot on the primary axis) since a larger speedup can be achieved at the cost of a larger error. On average (geometric mean), TUSQ performs $39.3\times$ and up to $3134.3\times$ faster than TQSim.

We also compare the performance of TQSim against TUSQ on smaller benchmarks (non-memory-critical), evaluated over 32 thousand shots (low compute), in Figure \ref{fig:comparison_with_tqsim} (B). For lower shots, the amount of redundancies (and the opportunity to eliminate them) is lower. This, coupled with the fact that the workloads are not memory-critical, makes this an unfavourable regime for TUSQ's operation. In this regime, TUSQ's pre-processing overheads do not give comparable returns. On average, TQSim is $3.26\times$ and $2.25\times$ faster than TUSQ for BV and QFT, respectively.  As the number of qubits grows (and the benchmarks become increasingly memory-critical), the speedup ratio decreases. This confirms our hypothesis that TUSQ is the optimal choice for time and memory critical benchmarks. We expect a similar behavior for other benchmarks as well.

\section{Discussion}\label{sec:discussion}

\subsection{Compatible Noise Models}\label{subsec:compatible_noise}
While DFTT+Caching can simulate arbitrary quantum channels, modelling noise as a unitary channel enables us to perform the simulation without requiring additional memory. This is true by definition for measurement and depolarizing noise. Many theoretical and simulation studies, especially those studying FTQC, model device noise as comprising only depolarizing and measurement error \cite{gidney2021stim, lin2024codesign, wang2021noise, garcia2024effects, viszlai2023matching, viszlai2025interleaved}. 

Decoherence (amplitude damping, phase damping, thermal relaxation, etc.) can also be implemented using Pauli channels using the twirling approximations given in \cite{ghosh2012surface}. The effect of decoherence can be expressed as:

\begin{equation}
{\scriptsize 
\rho \to (1 - p_X - p_Y - p_Z) \rho + p_X X \rho X + p_Y Y \rho Y + p_Z Z \rho Z
}
\label{eq:noise-characterisation}
\end{equation}

\ignore{\tina{
\begin{multline}
\rho \to (1 - p_X - p_Y - p_Z) \rho \\
+ p_X X \rho X + p_Y Y \rho Y + p_Z Z \rho Z
\label{eq:noise-characterisation}
\end{multline}}}

Here \( p_X = p_Y = \dfrac{1 - e^{-t / T_1}}{4} \), and \( p_Z = \dfrac{1 - e^{-t / T_2}}{2} - \dfrac{1 - e^{-t / T_1}}{4} \). The $p_X$ and $p_Y$ terms account only for amplitude damping errors ($T_{1}$ errors) while the $p_Z$ term accounts for both amplitude and phase damping errors ($T_{1}$ and $T_{2}$ errors.)

General non-unitary channels can be supported in the TUSQ framework by replacing DFTT with DFTT+Caching. The other components of TUSQ remain unchanged since they don't assume operator invertibility.

\subsection{Utility of TUSQ beyond NISQ}\label{subsec:tusq_utility} 

\begin{table}[h!]
    \centering
    \begin{tabular}{|c|c|c|c|}
    \hline
         Benchmark & QFT$\_40$ & Adder$\_40$ & QAOA$\_40$ $(p=2)$\\
         \hline
         Unopt TNS & $1119642$ & $628889$ & $158407$\\ 
        \hline
        TNS+TUSQ & $3444$ & $2625$ & $805$ \\
        \hline
    \end{tabular}
     \vspace{6pt}
    \caption{
    Time taken (in seconds) to perform an unoptimized Tensor Network Simulation (Unopt TNS) vs a TUSQ-enabled tensor network simulation (TNS+TUSQ). All benchmarks consist of 40 qubits and are evaluated for $\text{bond dimension} = 16$.}
    \label{tab:large_qubit_estimate}
\end{table}

TUSQ can be used to perform logical-level simulations of quantum programs with noise models characteristic of the Fault-Tolerant Quantum Computing (FTQC) era. Tools like Stim \cite{gidney2021stim} allow the user to perform physical qubit-level simulations with error correction. However, this is scalably possible for Clifford circuits only \cite{aaronson2004improved}. An error-corrected simulation of any useful quantum circuit at the physical qubit level is infeasible since they contain non-Clifford gates.

A practical strategy for a general non-Clifford circuit in the FTQC era is to obtain a logical-level noise model, corresponding logical error rates using Stim simulations, followed by a logical-level simulation. Since FTQC era applications are deep circuits with a large number of logical qubits, they end up being time and memory-critical on most systems. This warrants the need for simulators like TUSQ. 

\ignore{Most FTQC studies assume an infinite coherence budget and physical qubits. However, industry roadmaps \cite{GoogleFTQCRoadmap} project their final FTQC devices being composed of 10k-100k physical qubits and long but finite coherence times. Recent work \cite{dangwal2025variational} shows that for the FTQC era assumption of 10k physical qubits, we can implement variational algorithms with $\sim 27$ data qubits, and $\sim 14$ ancillae. While Ref.~\cite{dangwal2025variational} performs only stabilizer simulations for their evaluations, TUSQ would enable us to perform an exact noisy SVS of these circuits - the gold standard in this case. This facilitates better design decisions for these large circuits.}

Statevector simulations are also used in verifying subroutines used in the FTQC era. Magic State Cultivation (MSC) \cite{gidney2024magic} is a low-overhead technique for producing high-fidelity $T$ states. Ref.~\cite{gidney2024magic} uses statevector simulations to verify the correctness of MSC for code distances $d=3$ and uses a heuristic to conjecture its correctness for larger distances. Scalable statevector simulators will enable us to obtain verifiable results for larger code distances and speed up simulation for smaller cases. We used TUSQ, now incorporated with the ability to perform mid-circuit measurements, to simulate MSC circuits provided in \cite{gidney2024magic}. We compare it against the codebase provided by \cite{gidney2024magic} at \cite{gidney_2024_13777072}. Using the original code, we get a simulation time of $1166.69$ seconds for a $\text{code\_distance}=3$ MSC circuit (consisting of 18 qubits) for $p=10^{-4}$. TUSQ performs the same simulation in 2.24 seconds - a $520\times$ speedup. This demonstrates the usefulness of TUSQ for quickly verifying simulable sub-routines used extensively in the FTQC era.

Finally, although TUSQ's algorithm has been demonstrated primarily for noisy SVS in this work, it applies to any simulation paradigm which involves matrix-vector multiplications and sampling from a vector. For example, all of TUSQ's components can be applied on top of a Tensor Network Simulator to perform efficient noisy Tensor Network Simulations (TNS) of circuits with hundreds of qubits. Table \ref{tab:large_qubit_estimate} reports the time taken by unoptimized TNS vs TNS+TUSQ for three 40-qubit benchmarks - QFT, Adder, and QAOA ($p=2$) for $\alpha=0.01, \beta=100$ and 100,000 shots. For performing unoptimized TNS simulations, we use CUDA-Q with the ``tensornet-mps'' flag. All simulations were performed using $\text{bond dimension} = 16$. Note that while all TNS+TUSQ simulations completed within the 40-hour time limit imposed by Perlmutter, the unoptimized TNS did not complete within that time budget. The reported numbers were extrapolated after computing the time taken for $100, 1000$ and $10,000$ shots. As expected, the time increases linearly with the number of shots since these naive simulators do not look to eliminate redundancy. We observe an average speedup of $248.39\times$ provided by TUSQ+TNS over a naive TNS. This shows the applicability of TUSQ to simulation methods that scale beyond statevector simulations. 

A similar discussion is given in ref. \cite{patti2025augmenting} on the usability of redundancy-aware noisy simulation techniques to perform noisy SVS and large-scale noisy TNS.

\subsection{Related Work}\label{subsec:related_work}

An extensive amount of work has been done to improve noiseless circuit simulation speed and reduce memory overhead. Refs \cite{wu2019full, zulehner2018advanced} use sparsity and data compression to simulate more qubits. FlatDD \cite{jiang2024flatdd} and BQSim \cite{jiang2025bqsim} use decision diagrams to represent quantum circuits in a memory-efficient way. Hybrid simulators like HyQuas \cite{zhang2021hyquas} switch between multiple approaches to quantum circuit simulation based on patterns in the circuit. qHipster \cite{smelyanskiy2016qhipster} implements a distributed quantum circuit simulator. DM-Sim \cite{li2020density} proposes a way to perform efficient density matrix simulations.

There exist simulators which are especially optimized for subclasses of circuits. Stim \cite{gidney2021stim} is extensively used in error correction research to simulate stabilizer circuits in polynomial time. MatchCake \cite{gince2024fermionic, matchcake_Gince2023} is used to simulate matchgate circuits in polynomial time. Other application-specific simulators are - Refs \cite{huang2021logical, lykov2023fast, wang2023enabling}.

Finally, Refs \cite{wang2025accelerating, li2020eliminating, patti2025augmenting} propose alternate ways to speed up noisy circuit simulation. We have performed an extensive comparison against TQSim throughout our work.

\section{Conclusion}\label{sec:conclusion}
Noisy quantum circuit simulation, when implemented using an ensemble of circuits, ends up being substantially slower compared to a noiseless simulation. However, a large amount of this computation is redundant. In order to spot and eliminate these redundancies and improve simulation speed, we propose TUSQ. TUSQ has four components - ER Tallying, ER Commutation, Pruning, and Depth First Tree Traversal (DFTT). The first three components spot redundant or insignificant computation in the ensemble of circuits and eliminate them. Once all components of the ensemble are guaranteed to produce distinct outputs, we use DFTT to look for partial computational overlaps among these distinct components and reuse them across computations while expending no additional memory. Compared to Qiskit and CUDA-Q, TUSQ achieves an average speedup of $59.06\times$ and $13.38\times$, with a maximum value of $7878.03\times$ and $439.38\times$, respectively.
\section{Acknowledgement}\label{sec:acknowledgement}
This work is funded in part by the STAQ project under award NSF Phy-232580; in part by the US Department of Energy Office of Advanced Scientific Computing Research, Accelerated 
Research for Quantum Computing Program; and in part by the NSF Quantum Leap Challenge Institute for Hybrid Quantum Architectures and Networks (NSF Award 2016136), in part by the NSF National Virtual Quantum Laboratory program, in part based upon work supported by the U.S. Department of Energy, Office of Science, National Quantum 
Information Science Research Centers, and in part by the Army Research Office under Grant Number W911NF-23-1-0077. The views and conclusions contained in this document are those of the authors and should not be interpreted as representing the official policies, either expressed or implied, of the U.S. Government. The U.S. Government is authorized to reproduce and distribute reprints for Government purposes notwithstanding any copyright notation herein. 
FTC is the Chief Scientist for Quantum Software at Infleqtion and an advisor to Quantum Circuits, Inc. We also acknowledge the use of NVIDIA Quantum Cloud resources for preliminary experiments conducted in this research. We also thank Meng Wang for sharing TQSim’s GPU-compatible code, using which we executed some benchmarking experiments.

\bibliographystyle{IEEEtranS}
\bibliography{refs}

\end{document}